# Distributed Federated Learning-Based Deep Learning Model for Privacy MRI Brain Tumor Detection

Lisang Zhou*, Bazaarvoice Inc., Austin, 78759, TX, United States, Email: lzhou@berkeley.edu
Meng Wang, Newmark Group, San Jose, 95133, CA, United States, Email: wang070210@gmail.com
Ning Zhou, Zhejiang Future Technology LLC, Hangzhou, 311200, Zhejiang, China,  zhouning723@gmail.com
*Corresponding author

*Abstract*—Distributed training can facilitate the processing of large medical image datasets, and improve the accuracy and efficiency of disease diagnosis while protecting patient privacy, which is crucial for achieving efficient medical image analysis and accelerating medical research progress. This paper presents an innovative approach to medical image classification, leveraging Federated Learning (FL) to address the dual challenges of data privacy and efficient disease diagnosis. Traditional Centralized Machine Learning models, despite their widespread use in medical imaging for tasks such as disease diagnosis, raise significant privacy concerns due to the sensitive nature of patient data. As an alternative, FL emerges as a promising solution by allowing the training of a collective global model across local clients without centralizing the data, thus preserving privacy. Focusing on the application of FL in Magnetic Resonance Imaging (MRI) brain tumor detection, this study demonstrates the effectiveness of the Federated Learning framework coupled with EfficientNet-B0 and the FedAvg algorithm in enhancing both privacy and diagnostic accuracy. Through a meticulous selection of preprocessing methods, algorithms, and hyperparameters, and a comparative analysis of various Convolutional Neural Network (CNN) architectures, the research uncovers optimal strategies for image classification. The experimental results reveal that EfficientNet-B0 outperforms other models like ResNet in handling data heterogeneity and achieving higher accuracy and lower loss, highlighting the potential of FL in overcoming the limitations of traditional models. The study underscores the significance of addressing data heterogeneity and proposes further research directions for broadening the applicability of FL in medical image analysis.

## I. INTRODUCTION

Artificial Intelligence (AI) has achieved much progress in numerous domains, such as material properties prediction, sensor accuracy and other tasks [1, 2]. Thereinto, distributed training is incredibly important because it lets us work with huge datasets that would be too much for just one computer to handle [3]. This approach spreads the work across many computers, cutting down on the time it takes to teach complex AI models. It's also key for working together safely on AI projects, especially with methods like federated learning that protect privacy. Distributed training makes AI more accessible, allowing more people and organizations to take part in and gain from AI advances. It speeds up AI development and helps adapt AI to tackle a wide variety of challenges, from health care to driving cars without a driver, and even keeping an eye on the environment. Simply put, it's essential for making AI better in many tasks.

Machine learning technologies are now widely used across various sectors, including sentiment analysis, macromolecules classification, and autonomous driving, with the medical imaging sector attracting notable interest of late [4-6]. Traditionally, labeling medical images and diagnosing patient conditions was a time-consuming process for doctors and researchers. However, advancements in medical image classification technology have enabled more efficient and accurate disease diagnosis, facilitating the discovery of novel disease traits and underlying mechanisms by researchers. Therefore, these technological strides have significantly enhanced treatment approaches and patient survival rates.

At present, the medical imaging field primarily relies on a Centralized Machine Learning (ML) framework for developing image classification models. In this centralized approach, data is transferred to the cloud for the construction of the ML model. Users interact with this model via an Application Programming Interface (API), making requests to utilize one of the provided services. However, the privacy of patient image data, which is highly sensitive, is a crucial concern. In a centralized ML setup, transmitting sensitive data to a server poses a significant privacy breach risk. Alternatively, the Distributed On-Site Learning architecture, where the server dispatches the model to users for local training without model-to-model communication, does not adequately address this crucial task due to its inability to facilitate interaction among the individually trained models.

To solve the problem, Federated Learning (FL) can be considered as an effective solution [7, 8]. In a time when edge devices like smartphones, sensors, and vehicles are loaded with significant amounts of data, there's an increase in issues related to data privacy, limitations in network bandwidth, and the availability of these devices. Federated learning has come forward as a revolutionary approach, allowing local clients to work together in training a collective global model without having to centralize the data. This method tackles the problems associated with moving large amounts of data between edge devices and a central server by keeping the user data on the devices, thus safeguarding user privacy.

In the field of medical image analysis, Li et al. demonstrated how federated learning can enable the sharing of medical data across institutions without violating privacy regulations such as HIPAA [9], significantly enhancing the accuracy and efficiency of disease diagnosis. In the realm of autonomous driving technology, Pokhrel et al. utilized federated learning to facilitate model sharing between vehicles across different regions, accelerating the optimization of autonomous driving algorithms and improving their generalization capabilities [10]. Additionally, in the area of recommendation systems, Jie et al. used federated learning to protect user privacy while enhancing the personalization and accuracy of the recommendation systems [11]. These studies not only prove the advantages of federated learning in privacy protection but also showcase its broad application potential across various tasks. Despite considerable advancements, the application of FL in addressing privacy concerns in brain tumor diagnosis has not been extensively explored. This article seeks to employ the FL framework for training the brain tumor medical image dataset, selecting an appropriate algorithm and investigating optimal parameter values to achieve high test accuracy.

This paper is structured as follows: The section 2 details the selection process for preprocessing methods, FL algorithms, Convolutional Neural Networks (CNN) models, optimizers, and loss functions, including a thorough explanation of the implementation aspects. Subsequently, section 3 presents the experimental outcomes, providing an in-depth analysis of the effects of various approaches and the efficacy of distinct combinations, with the aim of identifying the optimal training approach. Finally, section 4 provides a comprehensive conclusion of this paper.

II. LITERATURE REVIEW

A. The Progresses of Federated Learning Algorithms

McMahan et al. established FedAvg as a foundational approach in Federated Learning (FL), where local models are averaged to update a global model [12]. Nonetheless, Karimireddy et al. observed that FedAvg tends to result in unstable and delayed convergence in varied environments. Li et al. further dissected the non-IID problem into components of systems heterogeneity and statistical heterogeneity [13]. Addressing this significant challenge, researchers have devised varied strategies across FL processes.

In the realm of local training, numerous effective strategies have been developed to regulate the local goal. Notably, Li et al. proposed adding a proximal term to bridge the divide between local and global models, and Acar et al. sought to synchronize these models through dynamic regularization [14]. Shi et al. approached the non-IID challenge by examining dimensional collapse [15]. Additionally, Qu et al. and Caldarola et al. investigated achieving flatter local minima through the use of a sharpness-aware optimizer [16].

On the server-side, enhancing the aggregation of local models and the updating process of the global model represent key strategies for overcoming the non-IID dilemma. Wang et al. introduced a normalized averaging method as an improvement over the straightforward aggregation of FedAvg [17]. Furthermore, Ma et al. performed local model aggregation on a layer-wise basis [18]. Conversely, Reddi et al. have shown that integrating momentum-based updates into the global model effectively counters local drift [19].

B. Deep Learning Algorithms on Brain Tumor

The application of diverse deep learning algorithms on brain tumor detection has been widely investigated in the past decade. For instance, Alqudah et al. designed a CNN methodology to categorize three types of brain tumors: glioma, meningioma, and pituitary tumor [20], achieving notable accuracies of 98.93% for cropped images, and 99% and 97.62% for uncropped and segmented images, respectively. Siar and Teshnehlab developed a CNN model capable of diagnosing brain tumors and multiple sclerosis (MS) simultaneously, attaining an accuracy rate of 96% [21]. Kang et al. applied transfer learning using pretrained deep CNNs to extract features from Magnetic Resonance Imaging (MRI) brain scans, demonstrating that leveraging a combination of deep features substantially boosts performance [22]. They observed that, especially with extensive datasets, the combination of a radial basis function with a support vector machine classifier yielded effective results. Qiu et al. proposed a transfer learning-based ResNet for brain tumor classification and detection, which achieved excellent performance [23].

Khan et al. introduced a system for identifying brain tumors using Deep Learning (DL) and advanced feature selection methods [24], achieving impressive accuracy rates of 97.8%, 96.9%, and 92.5% across various datasets. Amin et al. outlined a CNN framework for detecting brain tumors, which begins by enhancing MRI scans via a Discrete Wavelet Transform (DWT) fusion technique followed by the application of a partial diffusion filter to diminish noise [25]. Tumor detection was accomplished through global thresholding, and the processed images were subsequently categorized into specific tumor types using the CNN model. Their research, which employed five different brain tumor image datasets, showcased superior results when compared to previous blending techniques.

Sharif et al. developed a technique for the segmentation and classification of brain tumors through the use of active DL [26]. This approach began with the enhancement of the image's contrast, followed by the deployment of a saliency-based DL method to generate a saliency map. Subsequent thresholding and the refinement of the pretrained CNN model, Inception V3, were key steps. They also identified prominent features using rotated local binary patterns, optimizing the process with a particle swarm optimization technique. Their methodology was validated against the BRATS datasets from 2015, 2017, and 2018, showcasing notable accuracy improvements in tumor classification. Rehman et al. [27] introduced a model leveraging CNN architectures—AlexNet, GoogLeNet, and VGGNet—for the identification of brain tumors. They employed data augmentation on MRI images to expand the dataset size and reduce the likelihood of overfitting, with the VGG16 architecture reaching an accuracy of up to 98.69%. Ismael and Abdel-Qader [28] applied a multilayer perceptron neural network for the classification of brain tumor MRI images, utilizing statistical features for extraction. Their approach achieved a peak accuracy of 91.9%.

However, even though a variety of deep learning algorithms have been widely adopted for this task, there is a lack of research on the development of algorithms for patient data privacy. Federated learning, as a highly private algorithm compatible with deep learning models, can be considered for this task.

## III. METHOD

### A. Dataset Preparation

In this research, the MRI dataset comprises 3,260 T1-weighted contrast-enhanced images that have undergone processing and enhancement [29]. This dataset is organized into two main directories: Training and Testing, each of which is subdivided into four categories containing images of glioma tumors (803 images), meningioma tumors (905 images), pituitary tumors (814 images), and images without tumors (668 images), respectively.

Thereinto, glioma, meningioma, and pituitary tumors are three distinct types of brain tumors, each with unique characteristics and origins. Gliomas are a group of tumors that arise from glial cells, which are the supportive tissue of the brain. They vary in aggressiveness and can be either low-grade (slower growing) or high-grade (fast growing and more malignant). Meningiomas originate from the meninges, the protective membranes that cover the brain and spinal cord. They are most commonly benign and grow slowly, but their location can cause serious symptoms depending on the pressure they exert on the brain. Pituitary tumors develop in the pituitary gland, a small organ at the base of the brain that controls critical hormones affecting various body functions. These tumors are typically benign and can affect hormone levels, leading to a variety of symptoms. Each of these tumor types requires distinct diagnostic and treatment approaches due to their different behaviors and impacts on the brain and overall health. The images are presented in a grayscale color mode with a resolution of 512x512 pixels. Examples of these images can be seen in Figure 1.

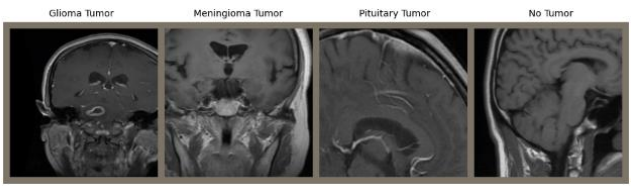

Figure 1. Sample images of brain tumor used.

To enhance classification accuracy, this study applied several preprocessing steps. Initially, due to the high resolution of the images, they were randomly cropped to a dimension of 224x224 pixels and converted to RGB color mode. To augment data variety, the images underwent horizontal flipping. Additionally, the conversion of images into PyTorch tensors involved normalizing pixel values from their original 0 to 255 integer range to floats between 0 and 1, along with adjusting the image dimensions to be compatible with PyTorch model architectures. The final step involved further normalization of the images to boost the model's performance and effectiveness. These preprocessing techniques collectively improved the model's ability to generalize and ensured greater consistency across the dataset.

### B. Preliminaries of Federated Learning

Federated Learning is a machine learning setting where the goal is to train a model across multiple decentralized devices or servers holding local data samples, without exchanging them. This approach is particularly beneficial for preserving privacy and reducing the need for data centralization and transmission. The basic idea involves training local models on individual devices and then aggregating these models to update a global model.

Let's denote the global model parameters by $w$, and assume there are N devices participating in the training process. Each device $i$ has a local dataset $D_i$ with $n_i$ samples. The objective of FL is to minimize the global loss function $L(w)$, which is typically a weighted sum of the local loss functions $L_i(w)$ computed on each device:

$$L(w) = \sum_{i=1}^{N} \frac{n_i}{N} L_i(w) \qquad (1)$$

During training, each device $i$ computes the gradients of its local loss function $L_i(w)$ with respect to the model parameters $w$, updates its local model, and then sends these updates to a central server. The server aggregates these updates, typically using an algorithm like Federated Averaging (FedAvg), to update the global model parameters. The FedAvg update can be represented as:

$$w_{\text{global}} = \sum_{i=1}^{N} \frac{n_i}{N} w_i \qquad (2)$$

Where $w_i$ represents the updated model parameters from device $i$. The updated global model is then distributed back to the devices for the next round of local training, and this process repeats for several iterations or until convergence.

### C. Convolutional Neural Network

Neural Networks are a class of machine learning models, highly effective for analyzing visual imagery [30-32]. Originating from the study of the brain's visual cortex, CNNs mimic the way humans perceive the world through layered processing of visual inputs. At their core, CNNs consist of multiple layers of convolutions interspersed with activation functions, pooling layers, and fully connected layers at the end. The convolutional layers apply filters to the input to create feature maps, highlighting different aspects of the input data, such as edges or textures. This operation reduces the spatial size of the representation and reduces the parameters in the network, making the neural network computationally efficient.

Unlike traditional neural networks, which require the manual selection of features before model training, CNNs automatically detect and learn the best features during the training process. This ability to learn hierarchical feature representations makes CNNs particularly powerful for tasks such as image and video recognition, image classification, and medical image analysis. Over the years, CNNs have been at the heart of major breakthroughs in computer vision, significantly surpassing previous benchmarks in accuracy and speed. Their architecture, capable of capturing the spatial and temporal dependencies in an image, enables them to make sense of the visual world in a way that is both deep and intuitive, opening up vast possibilities across various applications.

## D. EfficientNet-based Federated Learning Brain Tumor Classification

In this study, EfficientNet was chosen to combine with the federated learning algorithm. EfficientNet is a CNN architecture introduced by Google researchers in 2019 [33]. It's known for its efficiency in terms of computational resources and accuracy. The innovation behind EfficientNet is its scaling method, which systematically scales the network's width, depth, and resolution with a set of fixed scaling coefficients. This approach contrasts with previous practices where scaling was done more heuristically. Figure 2 presents the workflow of the proposed approach.

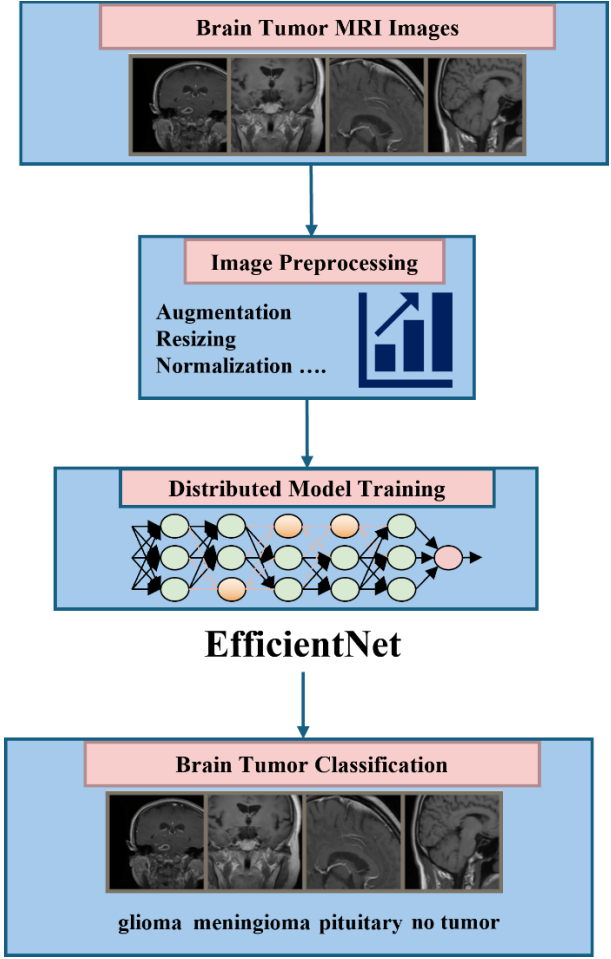

Figure 2. The workflow of the proposed approach.

EfficientNet uses a baseline model called EfficientNet-B0, developed through a neural architecture search that optimizes both accuracy and efficiency (FLOPS). The scaling method then expands this model into a family of EfficientNets (B1-B7) that provide a range of trade-offs for different application needs. This family of models has achieved state-of-the-art performance on benchmarks like ImageNet while being more efficient in terms of computing and parameters compared to previous models. The success of EfficientNet has made it widely adopted for various tasks in computer vision.

The research employed EfficientNet for data processing and feature detection in brain tumor image classification. Furthermore, it utilized FedAvg for the aggregation and averaging of each client's trained model, culminating in the development of a precise model. Figure 3 shows the mechanism diagram of the proposed federated learning-based method.

## E. Implementation Details

This research adjusted several hyperparameters, such as the total number of epochs, epochs per client, the total number of clients, the number of clients involved in each round of the global model update, the size of the mini-batches, and the learning rate. Regarding the optimization method, the research employed AdamW [34]. AdamW builds upon the adaptive learning rate features of Adam. It introduces weight decay regularization following the gradient calculation step, which enhances both the model's generalization ability and convergence rate.

Assuming that θ denotes the model weights, λ symbolizes the regularization coefficient, and η indicates the learning rate, the modification introduced by AdamW can be expressed with the addition of a momentum correction term, denoted as C.

$$\theta_t = \theta_{t-1} - \{\eta_t(C + \lambda\theta)_{t-1}\} \quad (3)$$

Regarding the choice of loss function, the research selected Cross-entropy loss. This loss function is prevalent in image classification tasks due to its unique characteristic of concentrating solely on the relevant category, eliminating the need for weight updates when the prediction is accurate. Cross-entropy serves to quantify the discrepancy between two probability distributions. In machine learning, if the actual probability distribution is represented by Y(X), and during training, an estimated distribution P(X) is used for approximation, then the Cross-entropy is calculated as follows: Assuming that θ denotes the model weights, λ symbolizes the regularization coefficient, and η indicates the learning rate, the modification introduced by AdamW can be expressed with the addition of a momentum correction term, denoted as C. Assuming that θ denotes the model weights, λ symbolizes the regularization coefficient, and η indicates the learning rate, the modification introduced by AdamW can be expressed with the addition of a momentum correction term, denoted as C.

$$H(Y,P) = -\sum_i^n Y(X = x_i) \log P(X = x_i) \quad (4)$$

For this image classification task, assuming there are n categories, a batch size of b, the actual distribution is denoted by Y, and the predicted distribution is represented as $\hat{Y}$, then the Cross-entropy loss formula is as follows:

$$LCE = -\frac{1}{b}\sum_i^b \sum_j^n y_{ij} \log \hat{y}_{ij} \quad (5)$$

## IV. RESULTS AND DISCUSSION

### A. The Performance of the Model

Figure 2 plotted the training curve against the number of global epochs ranging from 0 to 100. The left graph shows the accuracy of a model, starting very low and rapidly increasing, fluctuating between approximately 0.6 and 0.8, suggesting some variability in model performance as training progresses. The right graph depicts the loss, which starts high, at around 7, and shows a steep decrease, stabilizing close to 0 as the epochs increase, indicating the model is learning and improving its predictions over time. The peak accuracy was observed in the 70th global epoch, as depicted in Figure 4, achieving 83.12%, surpassing the 85% threshold. Meanwhile, the minimum loss recorded was 0.672.

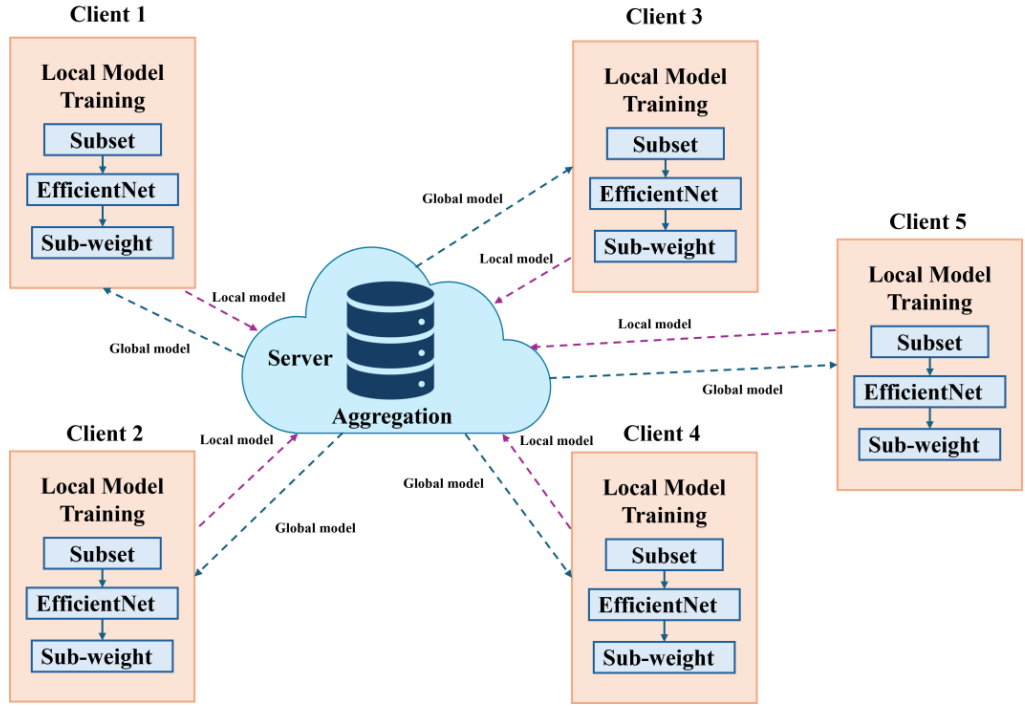

Figure 3. The mechanism diagram of the proposed federated learning-based method.

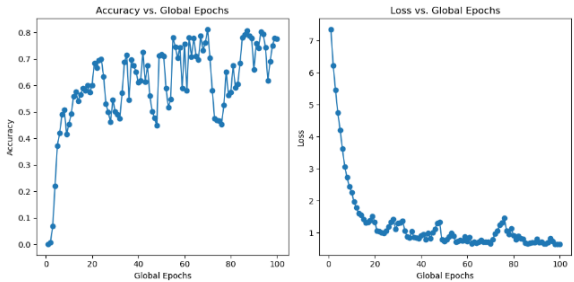

Figure 4. The performance of the EfficientNet model in terms of Accuracy and Loss.

Figure 5 illustrates the final training accuracy for each client following the concluding epoch of training. All clients achieved substantial training accuracies, with an average of 99%. Despite the data's variability leading to fluctuations in the curves, the trend in accuracy gradually ascends. To enhance accuracy further, incorporating strategies from algorithms such as FedProx and SCAFFOLD will be considered to mitigate data heterogeneity and bias effects. Additionally, this approach effectively minimizes the loss value, facilitating a quicker convergence of the loss curve.

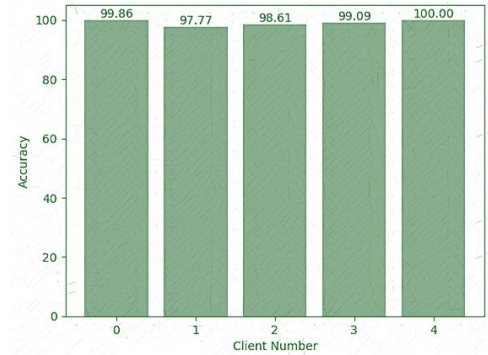

Figure 5. Accuracy per client after last global epoch.

### B. The Performance Comparison

The investigation into the performance of ResNet and EfficientNet via their accuracy and loss curves revealed that ResNet trails behind EfficientNet in terms of accuracy, loss mitigation, and curve smoothness shown in Table 1 and Figure 6. Notably, ResNet's performance is significantly hampered by data heterogeneity, leading to notable instability and insufficient convergence, even after 100 global epochs.

Table 1. Comparison of different CNN models using testing accuracy and loss

| CNN Model | Max Testing Accuracy | Min Testing Loss |
|---|---|---|
| EfficientNet-B0 | 80.17% | 0.612 |
| ResNet-50 | 65.32% | 1.017 |

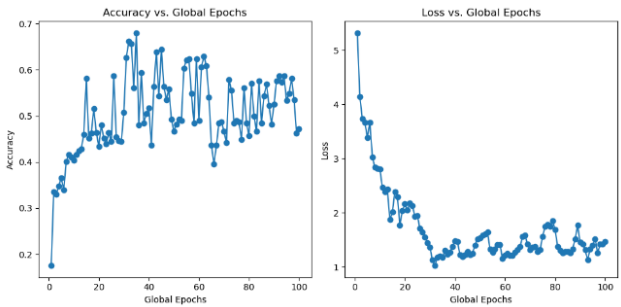

Figure 6. The performance of the ResNet model in terms of Accuracy and Loss.

ResNet, a convolutional neural network designed to deepen the model through deep residual learning, excels in capturing intricate data details. However, this depth comes at the cost of increased computational resources and extended training durations. In contrast, EfficientNet stands out for its use of Compound Model Scaling, which simultaneously fine-tunes the model's depth, width, and resolution. This capability ensures better adaptability to diverse data types and more effectively addresses the challenges of data heterogeneity. Additionally, EfficientNet leverages AutoAugment technology to enrich training with variedly manipulated images, enhancing its efficiency and accuracy while requiring fewer computational resources and shorter training times compared to ResNet.

For ResNet to match the performance benchmarks set by EfficientNet, this study suggests the necessity for advanced image preprocessing techniques and the allocation of more GPUs for training. This approach could potentially bridge the performance gap between ResNet and EfficientNet, optimizing ResNet's training efficiency and accuracy.

*C. Discussion*

The exploration of federated learning in our study reveals several key advantages that underscore its potential in overcoming prevalent challenges within the machine learning domain. Primarily, the adaptability and robustness of federated learning models are evident in their ability to handle data heterogeneity—an inherent characteristic of distributed data environments. This adaptability is crucial, as it enables the learning model to maintain high accuracy levels despite the variability and complexity of the data sources.

Furthermore, the experimental results highlight the effectiveness of federated learning in reducing model loss over time, which is indicative of the models' learning efficiency and their capability to improve prediction accuracy as more data is processed. This efficiency is not merely a reflection of algorithmic strength but also of the collaborative nature of federated learning, where diverse data points contribute to a more comprehensive learning process. Moreover, the achievement of substantial training accuracies across clients after the final training epoch, with an average of 99%, speaks volumes about the federated learning model's capability to unify diverse datasets towards achieving common learning objectives. This unification not only optimizes the learning process but also democratizes data, allowing for a more inclusive and equitable approach to machine learning.

Although this study has achieved encouraging results in the field of medical image recognition through the application of federated learning, there are still a series of challenges and directions for further development when applying federated learning more broadly to medical image analysis. First, the high heterogeneity of data and the differences in data distribution across different medical institutions may affect the model's generalization capability and accuracy. Moreover, while federated learning enhances data privacy, how to effectively manage and utilize medical data across institutions while ensuring data privacy remains a technical and ethical challenge. Additionally, complex medical image recognition tasks require models to possess high accuracy and interpretability, necessitating more sophisticated model designs and optimization methods. Some advanced models can be considered to further improve the performance of the models such as large language models, denoising autoencoder, reinforcement learning, data fusion, differential privacy [35-41]. In addition, the developed algorithms can be also combined with some advanced hardware technologies for real application [42, 43]. Future research could explore more advanced federated learning algorithms to improve model performance on multi-source heterogeneous data, while also paying attention to model interpretability to better enable medical professionals to understand and trust model predictions. Interdisciplinary collaboration will be key in advancing federated learning in the domain of medical image recognition, opening new research directions and application scenarios by combining medical expertise with artificial intelligence technology.

V. CONCLUSION

This research integrates Federated Learning with MRI datasets to enhance data privacy. By employing EfficientNet-B0 in conjunction with the FedAvg Algorithm, the investigation has formulated a classification method that is both adaptable and secure, outperforming recent approaches. Additionally, the study undertook a comparative analysis of various CNN architectures, highlighting the superior performance achieved through this novel combination.

Looking ahead, addressing data heterogeneity emerges as a significant challenge. Identifying effective strategies to boost accuracy amidst more diverse data sets stands as a crucial avenue for future research. Moreover, validating this method against other complex datasets will be essential to ascertain its robustness and applicability across different domains.